\documentclass[aps,prl,preprint,showpacs,preprintnumbers,amsmath,amssymb]{revtex4}
\usepackage{graphicx}
\usepackage{txfonts}

\begin{document}


\title{Interlayer magnetoresistance in an anisotropic pseudogap state}
\author{M. F. Smith}
 \email{mfsmith@physics.uq.edu.au}
\affiliation{%
Department of Physics, University of Queensland,4072 Brisbane, Queensland, Australia}
\author{Ross H. McKenzie}
\affiliation{%
Department of Physics, University of Queensland,4072 Brisbane, Queensland, Australia}
\date{\today}

\begin{abstract}
The interlayer magnetoresistance of a quasi-two-dimensional layered metal with a $d$-wave pseudogap is calculated semiclassically.  An expression for the interlayer resistivity as a function of the strength and direction of the magnetic field, the magnitude of the pseudogap, temperature, and scattering rate is obtained.    We find that the pseudogap, by introducing low-energy nodal quasiparticle contours, smooths the dependence on field direction in a manner characteristic of its anisotropy.  We thus propose that interlayer resistance measurements under a strong field of variable orientation can be used to fully characterize an anisotropic pseudogap.  The general result is applied to the case of a magnetic field parallel to the conducting layers using a model band structure appropriate for overdoped T$\ell$2201.

\end{abstract}


\maketitle
\section{Introduction}

High temperature superconducting cuprates, organic charge transfer salts, some heavy fermion materials and a host of other intriguing electronic systems, are layered metals in which electrons are approximately confined to a given atomic layer.  Much of the interesting behavior of these materials arise because of strong electronic correlations within a single layer.  Surprisingly, it turns out that one of the most effective means of accessing in-layer properties, particularly those properties that are highly anisotropic within a layer,  is to measure {\it interlayer} electronic transport coefficients in a strong magnetic field\cite{huss03,abde06,abde07,anal07,berg03,bali05,kart04,wosn}.

The interlayer electrical resistivity $\rho_{zz}$ depends on the direction of the magnetic field in a manner that is highly sensitive to the anisotropy of the quasi-2D band structure.  High-resolution maps of the Fermi surface, and other band structure properties, have already been obtained by fitting $\rho_{zz}$ data to calculations based on semiclassical magnetotransport theory.  This technique has been applied to a wide variety of layered materials including overdoped cuprates\cite{huss03,abde06,abde07,anal07}, ruthenates\cite{berg03,bali05}, and organic charge transfer salts\cite{kart04,wosn}. The $\rho_{zz}$ data also contains information about in-plane scattering and can be used to study the directional dependence of elastic and inelastic scattering rates\cite{sand01,kenn07,kenn08,kenn09,huss08,sing07,smit08}.  Notably, it has been used to reveal a $T$-linear, anisotropic scattering contribution in overdoped cuprate superconductors that appears to be tied to superconductivity itself \cite{abde06,tail06,fren09}.  It is important to press further, to ask what other anisotropic properties of the metallic layers can be detected and characterized via interlayer transport in high magnetic fields.

In this article we ask what interplane transport data can tell us about an anisotropic pseudogap $\Delta_{\bf k}$ in quasi-2D metals.  Since an anisotropic gap in the density of states will affect the field-direction dependence of $\rho_{zz}$, we expect that interlayer magnetoresistance can be used to map out $\Delta_{\bf k}$ as well.  A natural application of this technique would be to slightly overdoped cuprates.  For these materials, a model of a 2D metal with a small $d$-wave pseudogap (that is starting to emerge with reduced doping) is a plausible description of the metallic state at fields above $H_{C2}$ and semiclassical calculations of $\rho_{zz}$ may adequately capture transport properties.  To extract from $\rho_{zz}$ information about the doping, temperature and field dependence of $\Delta_{\bf k}$ would be of great value towards understanding the relationship between the pseudogap and superconductivity\cite{norm05,hufn08}.  The effects of a non-zero $\Delta_{\bf k}$ may already be present in existing interlayer resistance data on slightly overdoped cuprates, convoluted with the effects of anisotropic scattering\cite{husspc}.  If so, a reinterpretation of these data using models that incorporate a pseudogap could be fruitful.

We study a model with well-defined electronic quasiparticles existing in the presence of a $d$-wave pseudogap in the density of states.  The manner in which the opening of the pseudogap will change the interlayer resistivity is predicted and the following main results obtained:

    i) An expression for the interlayer resistance $\rho_{zz}$ in the semiclassical limit in a strong magnetic field of arbitrary strength and direction.

   ii) For the simple case of a field parallel to the layer, with arbitrary intralayer orientation $\phi_B$, the quantitative effect of a pseudogap on $\rho_{zz}(\phi_B)$ is calculated using a realistic model band structure.  The average magnitude of $\rho_{zz}(\phi_B)$ varies non-monotonically with the size of the pseudogap while its $\phi_B$ dependence is modified  in a manner distinctive of the pseudogap symmetry.  A strongly anisotropic normal-state $\rho_{zz}(\phi)$ is smoothed by the pseudogap through the introduction of new low-energy current contributions associated with $d$-wave nodes.

  Considering our results in light of the success of the AMRO technique in extracting band structure and scattering parameters of cuprates, we propose that this technique should also prove to be a viable means of obtaining a $T$-, $B$- and doping-dependent parametrization of the $d$-wave pseudogap.

\section{Semiclassical picture of Pseudogap State}

As a simple model of the $d$-wave pseudogap state one can use the normal (diagonal) part of the BCS Green's function, taking the anomalous part equal to zero.  The Green's function is
\begin{equation}
\label{gf}
G_0(\omega,{\bf k},x)=\bigg{(}\frac{u_{\bf k}^2}{\omega-E_{\bf k}}+\frac{v_{\bf k}^2}{\omega+E_{\bf k}}\bigg{)}
\end{equation}
with band energy $\xi_{\bf k}$, pseudogap $\Delta_{\bf k}$ and relative spectral weights for the electron and hole terms:
\begin{equation}
u_{\bf k}^2=\frac{1}{2}(1+\xi_{\bf k}/E_{\bf k})\;\;\; , \;\;\;v_{\bf k}^2=\frac{1}{2}(1-\xi_{\bf k}/E_{\bf k})
\end{equation}
and a quasiparticle energy $E_{\bf k}$ given by
\begin{equation}
E_{\bf k}=\sqrt{\xi_{\bf k}^2+\Delta_{\bf k}^2}.
\end{equation}

Recently Yang, Rice and Zhang (YRZ)\cite{yang06} proposed an ansatz for the coherent part of the Green's function in the pseudogap state of high-temperature superconducting cuprates based on renormalized mean field theory calculations of the resonating valence bond state in the $t-J$ model.  It has a similar form to Eq. \ref{gf}, differing only by the appearance of small additional terms in the band energy that break the particle-hole symmetry (also, YRZ proposed specific doping dependencies of the overall magnitude of the spectral weight of the coherent part, as well as of the band hopping parameters and $\Delta_0$).  YRZ found good agreement between their model and several characteristic features of ARPES data\cite{yang09}.  The interlayer resistance of the YRZ model could be studied using the same approach followed in this article.  We consider here the more familiar BCS expression (Eq. \ref{gf})  in order to illustrate the qualitative changes to the normal-metal $\rho_{zz}$ that are induced by turning on $\Delta_{\bf k}$.

Eq. \ref{gf} can be viewed as a description of a two-band metal with band energies of $\pm E_{\bf k}$, measured from the Fermi level, and ${\bf k}$-dependent spectral weights.  At $T=0$ the lower band is filled, the upper band empty and their nodal crossing point lies exactly at the chemical potential.  If the imaginary part of the self-energy correction to Eq. \ref{gf} is small (compared to relevant $\omega$) then the quasiparticles in each band are well-defined and transport properties can be calculated using a semiclassical Boltzmann approach.

For the semiclassical picture to be applicable, the quasiparticles in each band must remain well-defined, i.e. the imaginary parts of the self energy correction to Eq. \ref{gf} must be small compared to relevant frequencies $\omega$.  At low temperature and frequency, impurity scattering will dominate.  The associated scattering rate can be obtained following the procedure for d-wave superconductors\cite{allo09} and it is known that, at sufficiently low frequency the impurity scattering rate becomes larger than the frequency so the semiclassical picture of transport is not applicable.  At high temperature and frequency, strong inelastic scattering will also render the semiclassical approach invalid.  However there may exist an intermediate frequency range for which both the impurity and inelastic scattering rates are relatively small.  In this range, quasiparticles are sharply defined and the scattering rate $\tau^{-1}(\omega,{\bf k})$ can be evaluated at the quasiparticle pole $\omega=E_{\bf k}$.  We assume that such a frequency range exists and calculate the interlayer resistivity in a magnetic field using Boltzmann theory.

\section{Interlayer Resistance in the Pseudogap State in the Presence of an Arbitrary Magnetic Field}

To have interlayer current there must be a finite amplitude $t_\perp$ for hopping between adjacent layers.  However, according to Kennett and McKenzie\cite{kenn07}, the form of the interlayer conductivity does not depend on whether or not interlayer transport is coherent (i.e. it does not depend on the relative magnitude of $t_\perp$ and $\hbar\tau^{-1}$) as long as in-plane momentum is conserved during interlayer hopping.  We may thus carry out the calculation of the interlayer conductivity by supposing that a 3D Fermi (quasi-cylindrical) Fermi surface exists even when we are in the regime in which the Bloch vector in the interlayer direction $k_z$ is not well-defined.  Taking advantage of this, we simply add to $E_{\bf k}$ a term $-2t_\perp (k_x,k_y) \cos(k_zc)$ where $t_\perp (k_x, k_y)$ is the interlayer hopping coefficient and $c$ the distance between layers.  The associated interlayer velocity is $v_z(k_x,k_y,k_z)=2c\hbar^{-1}t_\perp(k_x,k_y)\sin(k_zc)$.  The calculation of the interlayer current is done to lowest order in $v_z$.

The Boltzmann equation in a weak electric field ${\bf \Xi}$ along the $z$ axis and a magnetic field ${\bf B}$ of arbitrary strength and direction is:

\begin{equation}
\label{beq}
\frac{\partial g_{\bf k}}{\partial t}-I[g_{\bf k}]=-e\Xi v_z({\bf k})\bigg{(}-\frac {df_0}{dE_{\bf k}}\bigg{)}
\end{equation}
where the total distribution is $f=f_0+f_1$ with $f_1=-(df_0/dE_{\bf k})g$, $f_0(x)$ is the Fermi function and $I[g]$ is the collision functional.  The auxillary time variable $t$ is defined by the equation of motion\cite{currnote}:
\begin{equation}
\label{eqmo}
\frac{d{\bf k}}{dt}=-e{\bf v}_g\times{\bf B}
\end{equation}
where ${\bf v}_g=dE_{\bf k}/d{\bf k}$.  Eqs. \ref{eqmo} and \ref{beq} are solved to obtain the distribution function, which is inserted into the expression for the interlayer current:
\begin{equation}
\label{jz}
j_z(t)=\frac{2e}{2\pi^3}\int d{\bf k} v_z({\bf k}[t])f_1(t,{\bf k}[t]).
\end{equation}
The current is found by taking a $t$-Fourier transform of $j_z(t)$ and evaluating in the zero-frequency limit.  The spectral weights from the two bands combine simply to give $u_{\bf k}^2+v_{\bf k}^2=1$ for the particle-hole symmetric case.

For a field ${\bf B}=B(\sin\theta_B\cos\phi_B,\sin\theta_B\sin\phi,\cos\theta_B)$, Eq. \ref{eqmo} gives $d\phi/dt=\omega_C(E,\phi,\theta_B)$ where the cyclotron frequency is
\begin{equation}
\hbar\omega_C(E,\phi,\theta_B)=eB\cos\theta_B\frac{{\bf v}_g\cdot{\bf k}_E}{k_E^2}.
\end{equation}
The cylindrical $\phi$ variable parameterizes the cyclotron orbit around a closed energy contour $E_{\bf k}=E$.  Any point ${\bf k}_E$ on the projection of this contour onto the $k_x-k_y$ plane is written as ${\bf k}_E=k_E(\phi)(\cos\phi,\sin\phi)$ where $k_E(\phi)$ is the radial cylindrical distance measured from some arbitrary point in the region enclosed by the contour.  In the normal state we can use a single energy contour (the Fermi surface) ${\bf k}_E={\bf k}_f=k_f(\phi)(\cos\phi,\sin\phi)$.  

The $k_z$ momentum varies according to
\begin{equation}
\frac{dk_z}{dt}=-\tan\theta_B\frac{d}{dt}\bigg{(}k_E(\phi)\cos(\phi-\phi_B)\bigg{)},
\end{equation}
which results in a periodic oscillation of the interlayer velocity $v_z(k_z[t])$ that is determined by the direction of the field angle $\theta_B$.  

Finally, since $E_{\bf k}$ is independent of $k_z$ in the collision functional, the integral over $k_z$ of the `scattering in' term vanishes by symmetry (this is true not only for scattering from point defects, but for any other scattering mechanism that can be regarded as spatially confined to a single plane\cite{smit08,kenn09}).  We are left with a relaxation time-description: $I[g_{\bf k}]=-g_{\bf k}/\tau(E_{\bf k})$ with the current relaxation rate equal to the total quasiparticle scattering rate $\tau^{-1}(\omega=E_{\bf k})$.  The fact that vertex corrections vanish to lowest order in $v_z$ in the calculation of the interlayer resistivity is a considerable simplification.  It means that we can use any appropriate model for the scattering rate, including elastic or inelastic scattering or even a sum over several different mechanisms.

We insert these expressions into Eq. \ref{beq}, formally solve for $g_{\bf k}(t)$ and use this in Eq. \ref{jz} to obtain the interlayer conductivity $\sigma_{zz}=1/\rho_{zz}$:
\begin{eqnarray}
\label{siggen}
\sigma_{zz}({\bf B})=A\int_{-\infty}^{+\infty}\frac{dE(-df_0/dE)}{1-P(E)}
\\
\times\int_0^{2\pi}\frac{d\phi_2t_\perp(\phi_2)}{\hbar\omega_C(\phi_2)}\int_{\phi_2-2\pi}^{\phi_2}\frac{d\phi_1t_\perp(\phi_1)}{\hbar\omega_C(\phi_1)}M(\phi_1,\phi_2)
\end{eqnarray}
where
\begin{equation}
\nonumber
A=\frac{e^2}{\hbar\pi^2}ceB\cos\theta_B,
\end{equation}
\begin{equation}
\nonumber
M(\phi_1,\phi_2)=G(\phi_1,\phi_2)\cos\Phi[\phi_1,\phi_2],
\end{equation}
\begin{equation}
\nonumber
G(\phi_1,\phi_2)=\exp(-\int_{\phi_1}^{\phi_2}\frac{d\phi^{\prime}}{\omega_C(\phi^{\prime})\tau(\phi)})
\end{equation}
\begin{equation}
\nonumber
\frac{\Phi(\phi_1,\phi_2)}{c\tan\theta_B}=[k_E(\phi_1)\cos(\phi_1-\phi_B)-k_E(\phi_2)\cos(\phi_2-\phi_B)],
\end{equation}
and $P=G(0,2\pi)$.  The function $G(\phi_1,\phi_2)$ is the probability that a quasiparticle can proceed from $\phi_1$ to $\phi_2$ along its cyclotron orbit without being scattered so $P$ is the probability that a quasiparticle completes an orbit.

Eq. \ref{siggen} has been written in the same form as the corresponding expression for a normal metal\cite{kenn07}.  However $\omega_C(\phi)$, $\tau$, $k_E(\phi)$ and $P$ all {\it depend on energy} in the pseudogap state (though we have not always written this explicitly).  Moreover, the $\phi$ and $k_E(\phi)$ variables must be interpreted differently in this expression depending on whether the energy is greater or less than $\max\Delta_{\bf k}$.  This is because these variables are defined with reference to a closed 2D cyclotron orbit but the orbits (i.e. the energy contours) have different topologies depending on the relative size of $E$ and $\max\Delta_{\bf k}$ as shown in Fig. \ref{Econ}.  For $E<\max\Delta_{\bf k}$ there are four equivalent (banana-shaped) contours closed around nodal points so a node can be taken as an orbit center with the polar angle $\phi$ parameterizing position along the contour.  Thus $k_E(\phi)$, which is measured from the node to the contour, depends strongly on both $E$ and $\phi$.  (We should include an overall sum over the four nodes in Eq. \ref{siggen}, though this has not been written explicitly.  There is no mixing of different nodes since an electron remains on a single nodal contour during cyclotron motion and the current contribution from each nodal region can be obtained separately.)  For $E>\max\Delta_{\bf k}$ a single contour encircles the entire normal state Fermi surface and $k_E(\phi)$, measured from a central point, is weakly anisotropic--i.e. its anisotropy is that of the normal state Fermi surface.  The energy integral in Eq. \ref{siggen} must be broken up into low and high energy regions with the $k_E(\phi)$ variable defined accordingly.

Eq. \ref{siggen} is the main result of this article.  This expression could be used in fitting procedures similar to those applied in the normal state of overdoped cuprates.  The magnitude of the pseudogap as, say, a function of doping, temperature and field strength in overdoped systems could then be extracted.  A typical set of fitting parameters might include hopping amplitudes describing the normal state band structure and interplane hopping magnitude $t_\perp$ (the values of which would be constrained by independent measurements and would be expected to be independent of temperature and weakly dependent on doping), the normal-state scattering rate $\tau^{-1}$ (which can also be independently estimated) and the gap magnitude.  Additional parameters could be incorporated if one were to go beyond the nearest neighbor expression for $\Delta_{\bf k}$, or to include anisotropy in the scattering rate.  Overall, the number of parameters would not have to exceed that used in previous normal state analysis.

We will not undertake detailed numerical evaluations of Eq. \ref{siggen} in this article but will discuss, in the remainder of this section, some of the general features of this expression that distinguish it from the familiar normal state result.  The contribution to the conductivity, Eq. \ref{siggen}, that comes from energies $E>>\max\Delta_{\bf k}$ will be identical to the normal state expression.  So, the total conductivity is a weighted sum of the normal state value and the low-energy (i.e. $E_{\bf k}<\max \Delta_{\bf k}$) contribution associated with the pseudogap.  The relative weighting is controlled by the value of $\Delta_0/k_BT$.  The properties of the low-energy (pseudogap) contribution to the conductivity are qualitatively different than those of the high-energy (normal state) contribution.  It is more strongly temperature dependent and is less sensitive to the direction of the magnetic field.

To get some feel for the low-energy contribution to the magnetoconductivity, associated with the pseudogap, we consider the nodal limit $E<<\Delta_0$ for which $\Delta_{\bf k}$ and $\xi_{\bf k}$ can be linearly expanded about nodal points.  In the nodal limit we write $\Delta_{\bf k}=v_2k_2$ and $\xi_{\bf k}=v_fk_1$ where $k_2$ and $k_1$ are momenta parallel and perpendicular to the Fermi surface, respectively.  The radius of the energy contour with energy $E$ is given by
\begin{equation}
\hbar k_E(E,\phi)=\frac{E}{\sqrt{v_f^2\cos^2(\phi-\phi_n)+v_2^2\sin^2(\phi-\phi_n)}}
\end{equation}
and the cyclotron frequency is
\begin{equation}
\omega_C(\phi,E)=eB\cos\theta_B E^{-1}\bigg{[}v_f^2\cos^2(\phi-\phi_n)+v_2^2\sin^2(\phi-\phi_n)\bigg{]}
\end{equation}
where $\phi_n$ is the direction of the node ($\phi_n=\pm \pi/4,\pm 3\pi/4$).  Since $v_f>>v_2$ the energy contour is a narrow ellipse and the cyclotron motion of the quasiparticle slows down dramatically as it crosses the Fermi surface $k_1=0$.

If the probability $P$ is small, then quasiparticles are unlikely to complete cyclotron orbits without being scattered and the field-dependence of the conductivity is weak.  The field-dependent effects of interest (i.e. the sensitivity of $\sigma_{zz}$ to in-layer anisotropy and the AMRO) occur when $P$ is of order 1.  The quantity $P$ depends on the scattering mechanism and, generally in the pseudogap state, on energy $E$.  In the simple case of point defects, the scattering rate\cite{allo09} is approximately given by:
\begin{equation}
 \tau^{-1}(E)=\tau^{-1}_0[\nu(E)/\nu_0]^{\eta}
 \end{equation}
where $\tau^{-1}_0$ is the normal state scattering rate, $\nu(E)$ and $\nu_0$ are the densities of states in the pseudogap and normal states, respectively, and $\eta=+1$ (or $-1$) in the Born (or unitary) limit.  For unitary scattering (to which we henceforth restrict ourselves) there is a cancelation in factors of the quasiparticle density of states so that $P$ becomes energy-independent.  In this case, $P$ has roughly the same value as in the normal state.  So, in strong fields, we can ignore the effects of scattering (i.e. set $G(\phi_1,\phi_2)=1$) in both the high-energy (normal state) contribution and the low-energy (nodal limit) contribution to $\sigma_{zz}$.  This simplifies the following discussion.

Both the sensitivity of $\sigma_{zz}$ to the anisotropy of the 2D band structure and the AMRO effect originate from the argument $\Phi(\phi_1,\phi_2)$ of the cosine in Eq. \ref{siggen}.  The cosine oscillates rapidly when $k_fc\tan\theta_B$ is large, and kills the integral everywhere except at special momentum directions, which depend on field orientation $\phi_B$.  As discussed in Ref. \onlinecite{kenn07}, the conductivity is thus dominated by the small region where both $\phi_1$ and $\phi_2$ are close to a special direction defined by the solution of
\begin{equation}
\label{specphi}
\frac{d}{d\phi}\bigg{[}k_f(\phi)\cos(\phi-\phi_B)\bigg{]}=0.
\end{equation}
Since the field direction $\phi_B$ determines the value of $\phi_1$ and $\phi_2$ that dominate the integrals, the band structure parameters are evaluated at a symmetry-unique point on the Fermi surface that can be tuned by field direction, allowing the Fermi surface to be mapped out.  Also, since the scale of the rapid oscillation is set by $k_fc\tan\theta_B$, the overall magnitude of the conductivity oscillates in $\theta_B$ with a period determined by this quantity (this is AMRO).

However, when we apply this reasoning to the low-energy pseudogap contribution, we find that such strong dependence on field direction angle is not expected.  The solution to Eq. \ref{specphi} in the nodal limit is $\phi-\phi_n=\arctan[\alpha^2(\phi_B-\phi_n)]$ where $\alpha=v_f/v_2>>1$.  The large factor $\alpha^2$ means that the dominant value of $\phi-\phi_n$ will almost always be close to $\pi/2$, i.e. close to the point at which the nodal energy contour crosses the Fermi surface, {\it independent} of the direction of field.  (The only exception would be if the magnetic field were pointed precisely in a nodal direction.)  So, the dependence on the field direction $\phi_B$ is far weaker in the low-energy pseudogap contribution than it is in the normal state.  Moreover, the scale for the oscillatory dependence in the nodal limit is $k_Ec\tan\theta_B\approx (E/v_2)c\tan\theta$.  For energies $E<<\Delta_0$ this quantity will be much smaller than one for any $\theta_B\neq \pi/2$.  The argument of the cosine in Eq. \ref{siggen} will be small and no oscillatory dependence on field angle $\theta_B$ will be seen.  Even at temperatures as high as $k_BT/\Delta_0\approx 1$ we do not expect to see prominent AMRO coming from the pseudogap contribution to the conductivity.  This is because the integral over energy will average $k_E$ over all values from $0$ to nearly $k_f$, giving no sharp period for oscillatory behavior.

These qualitative arguments suggest that the low-energy (pseudogap) contribution to the conductivity will not show the strong field-direction dependence characteristic of the high-energy (normal state) contribution.  (A detailed analysis is needed, however, to account for the strong energy-dependence of the scattering rate that could change this picture by giving dominant weight in the integral to a particular energy range.)  So, the onset of the pseudogap should have the generic effect of smoothing the dependence on field angle.  Nevertheless, this smoothing will proceed in a particular manner that is {\it characteristic of the anisotropy of the pseudogap}.  None of the above effects would occur for an isotropic pseudogap, and the $d$-wave case discussed here could be distinguished from alternative forms since the arrangement of nodal points would have a different relationship with the normal state band anisotropy.

In the next section we consider the simple limit of a field in the layers, i.e. $\theta_B=\pi/2$.  This is done to provide a more quantitative description of the effect that a pseudogap has  on the field-direction anisotropy of $\rho_{zz}(\phi_B)$.  Also, theoretical expressions for $\rho_{zz}$, with which we can compare our results, have been obtained previously using a different formalism.

\section{Case of a Field Parallel to the Layers}

The general result Eq. \ref{siggen} can be evaluated in the limit $\theta_B\to\pi/2$ (i.e., for the case of a field in the layers) by employing a stationary phase approximation but it is simpler to go back to the beginning of the derivation and make this assumption.  When ${\bf B}$ is in the layers:
\begin{equation}
\label{sigz}
\sigma_{zz}=\frac{e^2c}{\hbar\pi^2}\int d^2{\bf k} \bigg{(}-\frac{df_0}{dE_{\bf k}}\bigg{)}t_{\perp}^2({\bf k})\frac{\tau^{-1}(E_{\bf k})}{\tau^{-2}(E_{\bf k})+\Omega_C({\bf k})^2}
\end{equation}
where ${\bf k}$ is the momentum in the plane and
\begin{equation}
\Omega_C({\bf k})=ec|{\bf v}_g\times{\bf B}|.
\end{equation}
In the normal state, $\tau^{-1}(E)=\tau^{-1}$ and $\Omega_C({\bf k})=\Omega_C(\phi)$ are both independent of energy so the integral over $E_{\bf k}$ gives unity and Eq. \ref{sigz} reduces to a Fermi surface average.  The magnetic field becomes important when the $\phi$-averaged quantity $\Omega_C\approx ecv_fB$ becomes comparable to the scattering rate $1/\tau$.  Note that the criterion for field effects $\Omega_C\tau\gtrapprox 1$ is more favorable by a factor of $k_fc$ than the corresponding criterion for in-layer transport, where $k_fc\approx 10$ is typical in cuprates\cite{anal07}.

Eq. \ref{sigz} can also be obtained using a tunneling Hamiltonian approach, and a similar result was thus obtained in Ref. \onlinecite{bula99}.  The tunneling current is expressed as a convolution of spectral functions on adjacent layers. The gauge can be chosen such that the difference in the vector potential between adjacent layers is ${\bf A}=c(B_y,-B_x,0)$ and the corresponding spectral functions differ only by a momentum shift equal to $e{\bf A}$.  Evaluating the spectral functions in the quasiparticle approximation and using $\Omega_C\approx (E_{\bf{k}-e\bf{A}}-E_{\bf k})$ one obtains Eq. \ref{sigz}.  The advantage of the semiclassical approach followed here is that it can be generalized to describe fields out of the layers (Eq. \ref{siggen}).  In the remainder of this article we will, however, focus on the simple case of Eq. \ref{sigz}.  We go beyond the $k_BT<<\Delta_0$ nodal limit considered in Ref. \onlinecite{bula99} to consider arbitrary $k_BT/\Delta_0$ and a realistic normal-state band structure for cuprates in order to study the effect of a small pseudogap on the $\phi_B$ dependence of $\rho_{zz}$.
\begin{figure}
\begin{center}
\includegraphics[width=4.0 in, height=3.5 in]{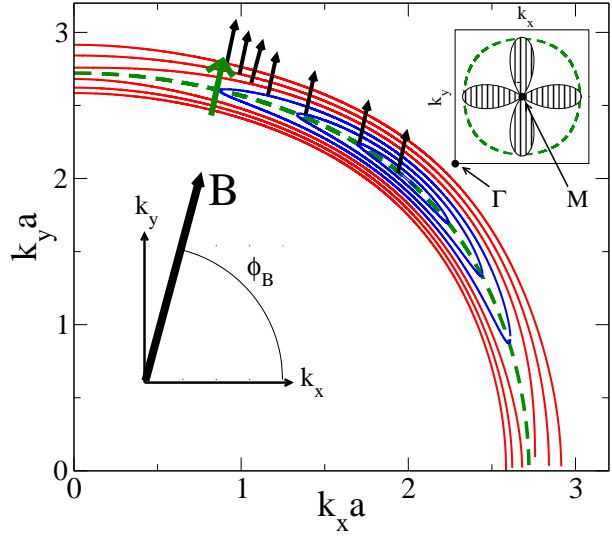}
\end{center}
\caption{\label{Econ} A small pseudogap reduces the dependence of $\rho_{zz}$ on the direction $\phi_B$ of a magnetic field ${\bf B}$ parallel to the layers.
Upper inset: The dashed (green) curve is a Fermi surface, closed around the corner $M$ point of the square Brillouin zone, and the hatched curve indicates the magnitude of the $d$-wave pseudogap.  Main panel: When the field is large, the interlayer current is dominated by ${\bf k}$-points on low-lying energy $E_{\bf k}$-contours at which the electron velocity ${\bf v}_g=dE_{\bf k}/d{\bf k}$ is parallel to ${\bf B}$.  The solid curves show low-lying $E_{\bf k}$-contours (moving outward from the node, the contours are for $E/\Delta_0=0.05,0.5,1,1.5,2,2.5,3$) in the upper right quadrant of the M-centred Brillouin zone. The arrows (each parallel to ${\bf B}$) are located at the dominant ${\bf k}$-point for each contour.  In the normal state, the dominant ${\bf k}$ is the point where the largest (green) arrow intersects the dashed (Green) Fermi surface.  In the pseudogap state, the dominant ${\bf k}$ are spread over a large range that extends from the normal state point (for $E_{\bf k}>\max\Delta_{\bf k}$) to the node (for $E_{\bf k}<<\max\Delta_{\bf k})$.  The opening of a pseudogap effectively spreads the current contribution over the Fermi surface, thereby smearing the $\phi_B$ dependence of $\rho_{zz}$.}   \end{figure}

In a strong magnetic field, $\Omega_C\tau>>1$ so $\Omega_C(\phi_B)\tau>>1$ at typical $\phi_B$, the Fermi surface average in Eq. \ref{sigz} is dominated by the ${\bf k}$ values for which $\Omega_C({\bf k})=0$, i.e. by ${\bf k}$ for which the quasiparticle velocity is parallel to ${\bf B}$.  This means that the normal-state interlayer resistivity is determined by the values of band parameters at a particular point on the Fermi surface ${\bf k}={\bf k}^*=k_f(\phi^*)(\cos\phi^*,\sin\phi^*)$ where the value of $\phi^*$ is controlled by $\phi_B$ (in an isotropic system $\phi_B=\phi^*$).  Moreover, the resistivity is independent of $\tau^{-1}$ in strong fields since the current is limited by classical magnetoresistance rather than scattering.  Upon varying $\phi_B$, one can use $\rho_{zz}$ to effectively map out the $\phi$-dependence of the in-plane band parameters.

In the pseudogap state, the energy dependence of $\Omega_C({\bf k})$ changes this simple picture, as illustrated in Fig. \ref{Econ}.  For energies $E_{\bf k}>>\max\Delta_{\bf k}$, the energy contours of the pseudogap state are almost identical to the Fermi surface itself.  So the contribution to $\rho_{zz}$ that comes from energies much larger than $\max \Delta_{\bf k}$ are the same as in the normal state.  However, when a pseudogap opens up (i.e. once $\Delta_0$ becomes comparable to $k_BT$) the conductivity begins to receive significant contributions from energies $E_{\bf k}<\max\Delta_{\bf k}$.  The associated low-energy energy contours are centered on nodes and the ${\bf k}$ point on such an energy contour where $\Omega_C({\bf k})$ vanishes is far removed from the corresponding normal state point ${\bf k}^*$.  This means that a small pseudogap results in contributions to $\rho_{zz}$ coming from a much broader range on the Fermi surface, thereby weakening the $\phi_B$ dependence.

\begin{figure}
\begin{center}
\includegraphics[width=4.0 in, height=3.5 in]{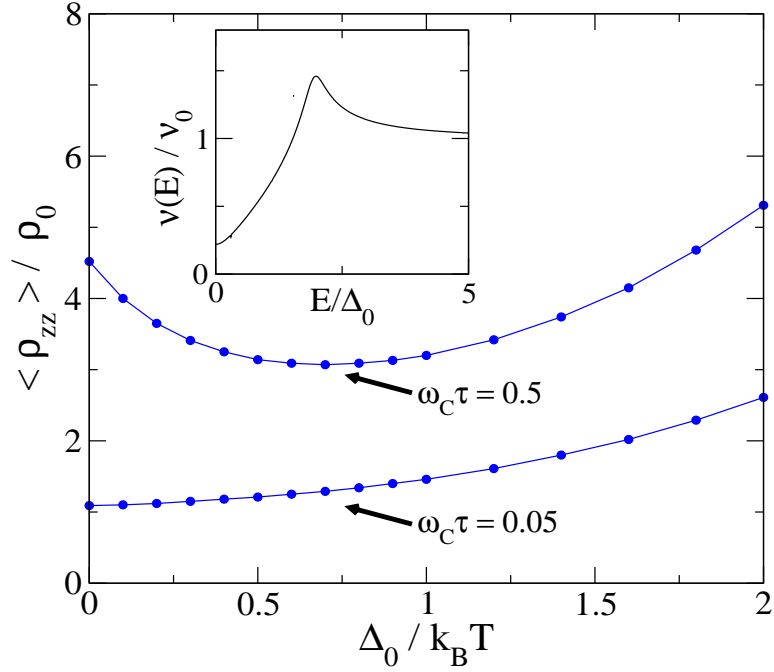}
\end{center}
\caption{\label{phiavg} The non-monotonic dependence of the interlayer resistivity $\rho_{zz}$ on the magnitude of a $d$-wave pseudogap.  Main panel: The vertical axis is the interlayer resistivity, averaged over the direction of the in-layer magnetic field $\phi_B$, in units of the normal state zero-field value $\rho_0$.  The horizontal axis is the magnitude of the $d$-wave pseudogap and the different curves are for different field strengths.  For weak fields ($\omega_C\tau<<1$), turning on the gap has no effect other than reducing the carrier density so the resistance increases with $\Delta_0/k_BT$.  In strong fields, the opening of a small gap reduces the average quasiparticle velocity and the associated Lorentz force responsible for the large magnetoresistance.  This effect results in an initial drop in the interlayer resistance.  As $\Delta_0/k_BT$ becomes large, the reduction of carrier density eventually overrides this effect and $\rho_{zz}$ begins to increase.  Upper inset: The pseudogap density of states $\nu(E)$ in terms of the normal state value $\nu_0$.  The scattering rate $\tau^{-1}$ depends on $E$ through the density of states.}   \end{figure}

This loss of $\phi_B$-dependence occurs initially without a corresponding increase in the magnitude of $\sigma_{zz}$.  In fact, the effect of turning on a small pseudogap in the presence of a strong in-layer magnetic field is to {\it decrease} the $\phi_B$ averaged interlayer resistivity, as shown in Fig. \ref{phiavg}.  This comes about because classical magnetoresistance is relieved by the pseudogap, both through a reduction of the average quasiparticle velocity and through the increased range of ${\bf k}$ that contribute to the current.  The effect is independent of $\tau^{-1}$ at large fields, as noted above, so the energy dependence of the scattering rate (i.e., whether we are in the Born or unitary limit) does not matter.  For a sufficiently large pseudogap, the reduction of the carrier density overcomes this effect, so $\rho_{zz}$ reaches a minimum at $\Delta_0/k_BT\approx 1$ and thereafter increases, eventually becoming very large for $\Delta_0/k_BT>>1$ when the current comes only from the nodal regions.

\section{Calculation of Interlayer Resistivity using Model Band Structure of T$\ell$2201}

To obtain a more quantitative picture of the $\phi_B$ dependence of $\rho_{zz}$ we use band structure parameters obtained from ARPES and interlayer resistance data on the two-layer cuprate T$\ell$2201.  The ARPES data\cite{plat05} can be reasonably fit by a tight binding model with nearest and next-nearest hopping parameters: $\xi_{\bf k}=-2t[\cos k_x +\cos k_y]-4t^{\prime}\cos k_x \cos k_y-\xi_0$ with ${\bf k}$ measured from $(\pi/a,\pi/a)$, $t'/t=0.42$ and $\xi_0/t=1.36$.  The resulting Fermi surface is shown in Fig. \ref{Econ}.  In this material the interlayer hopping parameter $t_\perp(k_x,k_y)$ vanishes by symmetry at 8 points on the Fermi surface (along $k_x=k_y$ and $k_x=0$ directions).  It can be modeled (according to AMRO data\cite{huss03}) as $t_\perp(\phi)=t_\perp[\sin 2\phi + k_{6}\sin 6\phi +(k_6-1.0)\sin 10\phi]$ with $k_{10}=k_6-1.0$ and $k_{6}=0.71$.  The energy scale $t_\perp$ can be absorbed into the zero-field, normal-state resistivity but the large anisotropy in $t_\perp(\phi)$ contributes to the strong $\phi_B$-dependence observed for this material in the normal state.  Moreover, since $t_\perp(\phi)$ vanishes at the nodes, the magnitude of $\rho_{zz}$ becomes extremely large in the nodal limit $\Delta_0/k_BT>>1$.
\begin{figure}
\begin{center}
\includegraphics[width=4.0 in, height=3.5 in]{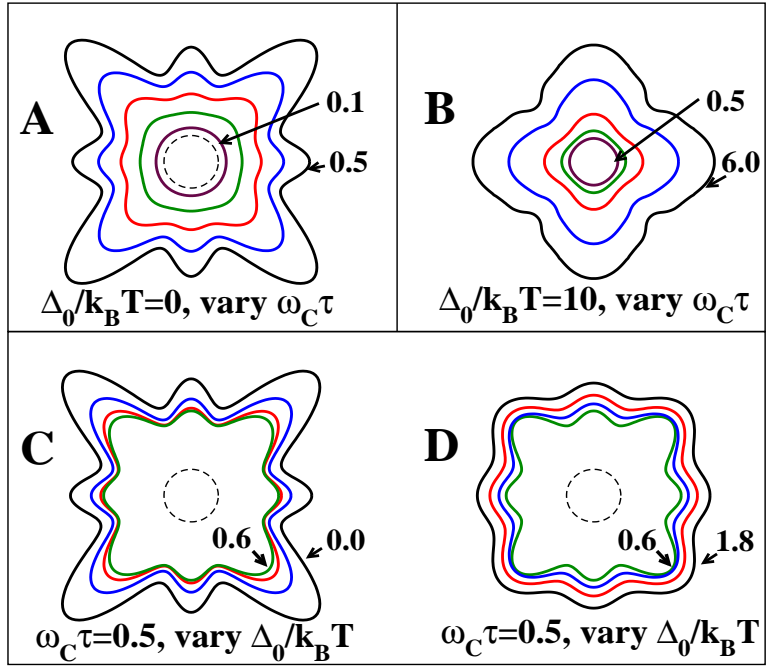}
\end{center}
\caption{\label{pol} Anisotropy of the interlayer resistance $\rho_{zz}$ in the normal and $d$-wave pseudogap states.  Solid curves are polar plots of $\rho_{zz}/\rho_0$ versus $\phi_B$ where a crystal axis is along the horizontal and the band structure of T$\ell$ 2201 has been used.  The dashed curve is the unit circle (unseen in panel B, where the radial scale is much larger).  Panel A: The normal state (constant $\Delta_0/k_BT=0)$ for varying field strength; the solid curves from inside out are for: $\omega_C\tau=0.1, 0.2, 0.3, 0.4, 0.5$.  Panel B: The low-temperature pseudogap state ($\Delta_0/k_BT=10)$ for varying field strength; the solid curves from inside out are for: $\omega_C\tau=0.5,1,2,4,6$.  Panel C and D: A large applied field (constant $\omega_C\tau=0.5$) with varying pseudogap magnitude.  The solid curves in C are, from outside in: $\Delta_0/k_BT=0,0.2,0.4,0.6$ and in D, from inside out: $\Delta_0/k_BT=0.6,1.0,1.4,1.8$.  The resistance first decreases (in C) then increases (in D) as the pseudogap grows.  The angle-dependence is reduced by the opening of a pseudogap and is eventually replaced by that associated with the anisotropic gap itself.}   \end{figure}

The anisotropic magnetoresistance in the normal state is illustrated in the polar plots of $\rho_{zz}(\phi_B)/\rho_0$ versus $\phi_B$ in Panel A of Fig. \ref{pol}.  A field of $\omega_C\tau\approx 0.5$ is sufficient to reveal the strong anisotropy of the underlying band structure.  Note also that the $\phi_B$-averaged magnitude of $\rho_{zz}/\rho_0$ decreases as the scattering rate increases for a given field strength.  In Panel B, the nodal limit $\Delta_0/k_BT>>1$ of $\rho_{zz}/\rho_{0}$ is depicted.  Here the current is coming entirely from momenta near the nodes and thus {\it provides no information about the normal state band parameters elsewhere on the Fermi surface}.  The anisotropy, which has been discussed in Ref. \onlinecite{bula99}, results from unequal, and $\phi_B$-dependent, contributions from different nodes owing to the large ratio of $v_f/v_2$ where $v_f$ is the Fermi velocity is the `gap' velocity.

Panels C and D of Fig. \ref{pol} describe the effect that turning on a $d$-wave pseudogap has on $\rho_{zz}(\phi_B)$ in a relatively strong field ($\omega_C\tau=0.5$).  The scattering rate was evaluated in the unitary limit, using the rounded density of states plotted in Fig. \ref{phiavg}.   In Panel C, which shows small values of $\Delta_0/k_BT$, the magnitude of $\rho_{zz}$ decreases as the gap opens.  In Panel D, which shows larger values of $\Delta_0/k_BT$, the $\phi_B$-averaged resistance has already reached its minimum value, depicted in Fig. \ref{phiavg} and is thus growing with $\Delta_0/k_BT$.

It is seen, by comparing Panels A and C, that the initial effect of a small gap on $\rho_{zz}(\phi_B)$ is similar to the effect of an enhancement in the scattering rate.  The reason for this similarity follows from the discussion of Fig. \ref{Econ}: the pseudogap increases the band of ${\bf k}$-points that contribute to the interlayer current just as would an increase in $\tau^{-1}$.  The manner by which the $\phi_B$ dependence changes as the pseudogap continues to grow in magnitude is, however, very different from that resulting from an increase in the scattering rate.  Not only does the magnitude of $\rho_{zz}$ vary non-monotonically with $\Delta_0/k_BT$, but $\rho_{zz}(\phi_B)$ evolves to incorporate the anisotropy of $\Delta_{\bf k}$ along with that already coming from $t_\perp(\phi)$ and the intra-layer band parameters.  Given the success of the angle-dependent interlayer resistance technique in extracting precise values for several band structure parameters, it appears that this technique should be equally capable of obtaining both the magnitude and anisotropy of a pseudogap as it emerges in the over- or near optimally doped cuprates.

\section{Conclusions}

Measurements of the interlayer resistivity in layered metals, made in a magnetic field with varying orientation, can be used to characterize anisotropic properties within individual layers.  Among these properties, band structure parameters and the inelastic scattering rate in various systems have already been extracted by this method.  In this article we have extended the analysis of such measurements to incorporate a pseudogap with $d$-wave symmetry.  A general expression for the interlayer resistivity in the pseudogap state was obtained via a semiclassical calculation.

For a field along the layers, the main effect of a small pseudogap is to smooth the dependence of the resistivity on the in-layer field direction $\phi_B$.  This occurs because, while electrons only contribute to the normal state interlayer current if they are located at a particular point on the Fermi surface, so that they have a velocity parallel to the magnetic field, quasiparticles with an energy smaller than the pseudogap can contribute to the interlayer current from anywhere on the Fermi surface. The average magnitude of the interlayer resistivity first decreases, then subsequently increases as the pseudogap opens, reaching a minimum value when the magnitude of the pseudogap is comparable to the temperature.  We hope that this work will stimulate new experiments and analysis to detect the presence and map the anisotropy of a pseudogap in layered strongly correlated materials.

We thank Matthew French and Nigel Hussey for helpful discussions and for sharing with us unpublished data.  This work has been supported by the Australian Research Council.

\end{document}